\documentstyle[aps,12pt]{revtex}

\begin{document}
\author{T. P. Cheng$^{*}$ and Ling-Fong Li$^{\dagger }$}
\address{$^{*}$Department of Physics and Astronomy, University of Missouri, St.
Louis, MO 63121\\
$^{\dagger }$Department of Physics, Carnegie Mellon University, Pittsburgh,
PA 15213}
\title{Why Naive Quark Model Can Yield a Good Account of the Baryon Magnetic Moments}
\date{CMU-HEP-97-10, DOE-ER-40682-135, hep-ph/9709xxx}
\maketitle

\begin{abstract}
The chiral quark model suggests that the baryon quark-sea is negatively
polarized. This modifies the spin structure as given by the naive quark
model and agrees with experimental data. However, for the magnetic moments,
there is significant cancellation between the contritutions from this sea
spin-polarization and the orbital angular momentum so that effectively the
moments are given by the valence constituent quarks alone, as in the NQM.
\end{abstract}

\medskip

Ever since the discovery\cite{emc} that the proton spin content is very
different from that given by the naive quark model (NQM), one of the puzzles
has been: why is the same naive quark spin structure capable of giving such
a good account of the baryon magnetic moments? In this paper we shall
suggest, in the context of the chiral quark model ($\chi QM$), a qualitative
explanation.

The basic idea of $\chi $QM\cite{mgtheor} is that the nonperturbative QCD
phenomenon of chiral symmetry breaking ($\chi $SB) takes place at distance
scale significantly smaller that of color confinement. Thus in the interior
of a hadron, but not so small a distance that perturbative QCD is
applicable, the effective degrees of freedom are the constituent quarks and
the $\chi $SB Goldstone bosons (GBs). Prior chiral quark model study has
indicated that the various nucleon flavor and spin puzzles can be understood
by the presence of a quark sea which is perturbatively generated by valence
quark's emissions of internal GBs\cite{ehq}\cite{CL95}\cite{CLsch97}\cite
{BrodMa}. This model can naturally account for the $\bar{u}$-$\bar{d}$
asymmetry as measured by the deviation from the Gottfried sum rule\cite{nmc}
and by the Drell-Yan processes\cite{na51}, as well as a strange quark
content consistent with the various phenomenological determinations\cite
{CL97}. The axial coupling of GBs and constituent quarks can modify the spin
content because the GB emission by a valence quark flips the quark spin
direction: 
\begin{equation}
q_{\pm }\longrightarrow q_{\mp }^{\prime }+GB\longrightarrow q_{\mp
}^{\prime }+\left( \bar{q}^{\prime }q\right) _{0}.  \label{GBemit}
\end{equation}
The subscripts denote the helicity states. We shall call the three quarks
(in $S$-wave state) of the NQM as the {\em valence quarks} and all the other
quarks (and antiquarks) broadly as the {\em quark sea}. The processes in (%
\ref{GBemit}) lead to a quark sea $\left( q^{\prime }\bar{q}^{\prime
}q\right) $ which is polarized (as given by $q_{\mp }^{\prime }$) in the
opposite direction to the baryon spin. (At the leading perturbative order,
the antiquark $\bar{q}^{\prime }$ and $q$ in the sea are not polarized
because they are produced through the spin-zero GB channels\cite{CL96}\cite
{smc-semi}.) In this way, we find that the quark contribution to the baryon
spin is substantially reduced from that of the NQM, in agreement with the
phenomenological result obtained by several generations of deep inelastic
polarized lepton-nucleon scattering experiments\cite{emc}\cite{smc}.

This reduction of the quark polarization also implies a significant decrease
of the quark-spin contribution to baryon magnetic moments. It is then
puzzling why the original quark model (without a polarized quark-sea) can
yield such a good description of the magnetic moments. Our $\chi QM$
explanation is that the quark sea must also carry a significant amount of
orbital angular momentum. In fact, angular momentum conservation implies
that the final state quark $q^{\prime }$ and $\left( \bar{q}^{\prime
}q\right) $ in the GB emission process (\ref{GBemit}) must be in a relative $%
P$-wave state. This orbital angular momentum, which is parallel to the
baryon spin, makes a positive contribution to the baryon magnetic moment and
thus compensates the quark-spin's reduction.

When we separate the spin and the orbital angular momentum contributions, we
are using the nonrelativistic approximation, which can provide us with an
intuitive physical picture of the hadron structure. As we shall comment on
at the end of the paper, existent chiral quark field theory calculations
also support our explanation.

From the SU(6) wavefunction of NQM we can calculate the number of valence
quarks with polarization $\sigma _{Z}=\pm 1\;$(denoted by particle names
with subscript $\pm $). In the case of the proton with valence quarks $%
\left( uud\right) $, we have 
\begin{equation}
u_{v+}=\frac{5}{3}\;\;\;\;u_{v-}=\frac{1}{3}\;\;\;\;d_{v+}=\frac{1}{3}%
\;\;\;\;d_{v-}=\frac{2}{3}.  \label{val-pol}
\end{equation}
The quark contribution to the baryon spin being the sum of the quark and
antiquark polarizations $\Delta q=\Delta _{q}+\Delta _{\bar{q}}=\left(
q_{+}-q_{-}\right) +\left( \bar{q}_{+}-\bar{q}_{-}\right) ,$ and because
there is no antiquarks and strange valence quark, we have 
\begin{equation}
\Delta u_{v}=\frac{4}{3}\;\;\;\;\;\Delta d_{v}=-\frac{1}{3}\;\;\;\;\;\Delta
s_{v}=0\;  \label{NQMspin}
\end{equation}
which makes up the total proton spin,$\;\Delta \Sigma _{v}=\Delta
u_{v}+\Delta d_{v}+\Delta s_{v}=1.\;$When it comes to the quark spin
contribution to the baryon magnetic moment,$\;\;\mu \left( B\right)
=\sum_{q}\left( \tilde{\Delta}q\right) _{B}\mu _{q}\;\;$with $\tilde{\Delta}q
$ $=\Delta _{q}-\Delta _{\bar{q}}$ (as antiquarks have opposite charges)$.\;$%
In the NQM with $\bar{q}_{+}=\bar{q}_{-}=0$ (thus $\tilde{\Delta}%
q_{v}=\Delta q_{v}$), we have, from Eq.(\ref{NQMspin}): 
\begin{equation}
\mu \left( p\right) _{v}=\frac{4}{3}\mu _{u}-\frac{1}{3}\mu _{d}=\left( 
\frac{e}{2M}\right)   \label{magmo-v}
\end{equation}
where we have used $\mu _{u}=-2\mu _{d}=e/3M$ reflecting the mass relation
of $M_{u}=M_{d}\equiv M.$  The results for octet baryons yield a good
account of the measured moments with $\mu _{d}\simeq -0.9\,n.m.$ (nucleon
magneton) corresponding to a set of constituent quark mass values close to
those used in fitting other hadron properties\cite{asym-mom}.

Ever since the publication by EMC of their experimental result\cite{emc}, it
is known that the proton spin content is quite different from that given by
the NQM of Eq.(\ref{NQMspin}), 
\begin{eqnarray}
\Delta u_{\text{expt}} &=&\;\,0.82\pm 0.06\;\;\;\;\;\;\;\Delta d_{\text{expt}%
}=-0.44\pm 0.06\;\;  \nonumber \\
\Delta s_{\text{expt}} &=&-0.11\pm 0.06\;\;\;\;\;\;\;\Delta \Sigma _{\text{%
expt}}=\;\,0.27\pm 0.11,  \label{expt-spin}
\end{eqnarray}
showing clearly that a good portion of the proton spin arises from something
other than quark spins\cite{ellisk}.

Besides the problem of understanding why the valence quarks can give by
themselves a good account of the baryon magnetic moments, we have another
related puzzle. Suppose we make the $ad\;hoc$ assumptions that the magnetic
structure is still given entirely by the quark spins and that antiquarks are
not polarized, as done in Ref.\cite{Karl}, $\mu \left( B\right)
=\sum_{q}\left( \Delta q\right) _{B}\mu _{q}.\;$Even though the baryon spin
content is significantly different from that given by the valence quarks: $%
\Delta q_{\text{expt}}\neq \Delta q_{v}$, one finds that $\Delta q_{\text{%
expt}}\;$can also lead to a good description of $\mu \left( B\right) .$
Namely, somehow, we get $\sum_{q}\Delta q_{v}\mu _{q}\simeq \sum_{q}\Delta
q_{\text{expt}}\mu _{q}^{\prime }.\;$This however requires a $\mu
_{d}^{\prime }\simeq 1.4\,n.m.$--- an approximately 50\% shift of the
effective quark moments and masses. Thus we have the puzzle that in some way
both $\Delta q_{v}$ and $\Delta q_{\text{expt}}$ can yield a good account of
the baryon magnetic moments. But, only for (the phenomenologically
incorrect) $\Delta q_{v}$ the fit leads to a set of correct quark masses.

We now discuss the $\chi $QM resolution of these puzzles. As explained in
the introduction, we need to calculate the spin and magnetic moment
contributions by the quark sea as generated by the internal GB emission
processes of the type in (\ref{GBemit}). We shall be working, for
simplicity, in a $\chi $QM with a flavor-$U(3)$ symmetry broken down to $%
SU(3)\times U\left( 1\right) $: the quark and GB form degenerate multiplets,
but with distinctive couplings for the octet GBs and the singlet $\eta
^{\prime }$ meson: $g_{1}/g_{8}\equiv \zeta \neq 1$. (In fact from our prior
study\cite{CL95} we expect $\zeta \simeq -1$ in this symmetric limit.) The
transition probability for the process of\ $q_{\pm }\longrightarrow q_{\mp
}^{\prime }+GB$ is parametrized to be 
\begin{eqnarray}
P(u &\rightarrow &d+\pi ^{+})=P(u\rightarrow s+K^{+})=a  \nonumber \\
P(u &\rightarrow &u+\pi ^{0})+P(u\rightarrow u+\eta )+P(u\rightarrow u+\eta
^{\prime })=\frac{1}{3}\left( 2+\zeta ^{2}\right) a.  \label{prob}
\end{eqnarray}
For any initial state $q,$ the total transition probability for $\left(
q\rightarrow all\right) $ is simply 
\begin{equation}
P\left( q\right) =\frac{1}{3}\left( 8+\zeta ^{2}\right) a.\;  \label{totprob}
\end{equation}

All calculations of the various angular momentum and magnetic moment
contents of the quark sea involve a ``three-part convolution'': the
contributions by a single reaction are to be multiplied by the transition
probability of the reaction Eq.(\ref{prob}) and by the number of initial
valence quarks of Eq.(\ref{val-pol}). To calculate the spin polarization of
the sea, the quantity for an individual process in this convolution involves
a count of $\pm 1$ (in units of $\frac12 \hbar $) for the two helicity
states multiplied by $\pm 1$ for the creation or destruction of a particular
quark flavor, etc. Keeping in mind that $\Delta _{\bar{q}}=0$ because to
this order $\bar{q}_{+}=\bar{q}_{-}$ in the sea, we obtain 
\begin{equation}
\Delta u_{sea}=-\frac{37+8\zeta ^{2}}{9}a,\;\;\;\;\;\;\;\Delta d_{sea}=-%
\frac{2-2\zeta ^{2}}{9}a,\;\;\;\;\;\;\;\Delta s_{sea}=-a.  \label{sea-pol}
\end{equation}
Their sum is the total spin polarization of the quark sea: 
\begin{equation}
\Delta \Sigma _{sea}=-\frac{2}{3}\left( 8+\zeta ^{2}\right) a=\Delta \sigma
\cdot P\left( q\right) \cdot \Delta \Sigma _{v}.  \label{seapol-tot}
\end{equation}
Namely, it is the product of the helicity-change per reaction regardless of
quark flavor $\Delta \sigma =-2$, the total transition probability Eq.(\ref
{totprob}) and the number of initial valence quarks weighted by the spin
directions (hence effectively the total valence quark polarization $\Delta
\Sigma _{v}=1$). By taking parameters such as $a\simeq 0.1$ and $\zeta
\simeq -1$ one can then get a fair account\cite{CL95} of the observed spin
structure (\ref{expt-spin}). This includes the reduction of the nucleon
axial vector coupling $g_{A}$ from $5/3$ to around $1.2$. All these changes
from the NQM values are interpreted as the renormalization effects due to
the quark sea.

The sea quark spin contribution to the proton magnetic moment is given by 
\begin{eqnarray}
\mu \left( p\right) _{spin} &=&\Delta u_{sea}\mu _{u}+\Delta d_{sea}\mu
_{d}+\Delta s_{sea}\mu _{s}  \nonumber \\
&=&-\frac{7+2\zeta ^{2}}{3}a\left( \frac{e}{2M}\right) \equiv \kappa
_{spin}\left( \frac{e}{2M}\right)  \label{magmo-sea}
\end{eqnarray}
It is easy to check that for octet baryons in general, because of the SU(3)
symmetric nature of the calculation, we have $\mu \left( B\right)
_{spin}=\kappa _{spin}\mu \left( B\right) _{v}$. This explains why $\mu
\left( B\right) =\mu \left( B\right) _{v}+\mu \left( B\right) _{spin}=\left(
1+\kappa _{spin}\right) \mu \left( B\right) _{v}$ can be fitted with $\Delta
q_{\text{expt}}$ by a simple rescaling of the effective quark moments as $%
\Delta q_{\text{expt}}\simeq \Delta q_{v}+\Delta q_{sea}$.

This change of angular momentum $\Delta \sigma \cdot \frac{1}{2}=-1\;$due to
quark spin flip in reaction (\ref{GBemit}) must be compensated by a
final-state orbital angular momentum. We shall describe this orbital motion
of the $\chi QM$ quark sea as due to the rotational motion of the two bodies
in (\ref{GBemit}). In their center-of-mass frame ($i.e.$ the rest frame of
the initial valence-quark), the orbital angular momentum is simply given by $%
{\bf l=r\times p\;}$where ${\bf r}$ and ${\bf p}$ are the relative
displacement and momentum vectors: ${\bf r=r}_{1}-{\bf r}_{2},\;{\bf p=p}%
_{1}=-{\bf p}_{2}$ with ${\bf r}_{1}=\frac{m_{2}}{m_{1}+m_{2}}{\bf r},$ $etc.
$ The hadronic matrix element of this operator can be evaluated, even
without the explicit knowledge of the baryon wavefunction, because angular
momentum conservation requires that 
\begin{equation}
\left\langle l_{Z}\right\rangle =1  \label{L=1}
\end{equation}
so as to compensate the quark spin change$.$ The total orbital angular
momentum of the sea can be calculated in the same way as the total spin
polarization of Eq.(\ref{seapol-tot}): 
\begin{equation}
\left\langle L_{Z}\right\rangle =\left\langle l_{Z}\right\rangle \cdot
P\left( q\right) \cdot \Delta \Sigma _{v}=\frac{1}{3}\left( 8+\zeta
^{2}\right) a.  \label{orb-sea}
\end{equation}
Thus, according to $\chi QM,$ the proton spin is built up from quark spin $%
\Delta \Sigma =\Delta \Sigma _{v}+\Delta \Sigma _{sea}$ and orbital angular
momentum: 
\begin{equation}
\frac{1}{2}\Delta \Sigma +\left\langle L_{Z}\right\rangle =\frac{1}{2}.
\end{equation}
Namely, the NQM spin sum $\Delta \Sigma _{v}=1$ is redistributed from the
valence quarks to the spin and orbital angular momenta of the quark sea: $%
\Delta \Sigma _{sea}$ and $\left\langle L_{Z}\right\rangle ,$ which is
constrained by the angular momentum conservation condition: 
\begin{equation}
\frac{1}{2}\Delta \Sigma _{sea}+\left\langle L_{Z}\right\rangle =0,
\label{angmo-cons}
\end{equation}
as seen in Eqs.(\ref{orb-sea}) and (\ref{seapol-tot}).

We now perform the three-part calculation of the orbital angular momentum
contribution to the magnetic moment. The orbital moment of each process $\mu
\left( q_{\pm }\rightarrow q_{\mp }^{\prime }+GB\right) $ is$:$%
\begin{equation}
\mu \left( q_{+}\rightarrow q_{-}^{\prime }\right) _{L}=\frac{e_{q^{\prime }}%
}{2M}\left\langle l_{qZ}\right\rangle +\frac{e_{q}-e_{q^{\prime }}}{2\tilde{m%
}}\left\langle l_{GB,Z}\right\rangle  \label{mu-single}
\end{equation}
where $\left( l_{q},l_{GB}\right) $ and $\left( M,\,\tilde{m}\right) $ are
the orbital angular momenta and masses of quark and GB, respectively. The
one unit of angular momentum in (\ref{L=1}) is shared by the two bodies: 
\begin{equation}
\left\langle l_{qZ}\right\rangle =\frac{\tilde{m}}{M+\tilde{m}}\;\;\;\;\text{%
and}\;\;\;\;\;\;\left\langle l_{GB,Z}\right\rangle =\frac{M}{M+\tilde{m}}.
\label{L1L2}
\end{equation}
The result (\ref{mu-single}) is then multiplied by the probability for such
a process to take place, to yield the magnetic moment due to all the
transitions starting with a given valence quark$:$%
\begin{eqnarray}
\left[ \mu \left( q_{\pm }\rightarrow \right) _{L}\right] &=&\pm \left[ \mu
\left( q_{+}\rightarrow q_{-}^{\prime }\right) _{L}+\mu \left(
q_{+}\rightarrow q_{-}^{\prime \prime }\right) _{L}+\frac{2+\zeta ^{2}}{3}%
\mu \left( q_{+}\rightarrow q_{-}\right) _{L}\right] a  \nonumber \\
&=&\pm \frac{9M^{2}+\left( \zeta ^{2}-1\right) \tilde{m}^{2}}{3\tilde{m}%
\left( M+\tilde{m}\right) }a\left( \frac{e_{q}}{2M}\right) .
\end{eqnarray}
The last step is to multiply the valence-quark-numbers, Eq.(\ref{val-pol}).
Thus for a baryons $B=\left( q_{1}q_{1}q_{2}\right) $ we have $\mu \left(
B\right) _{orbit}=\frac{4}{3}\left[ \mu \left( q_{1+}\rightarrow \right)
_{L}\right] -\frac{1}{3}\left[ \mu \left( q_{2+}\rightarrow \right)
_{L}\right] .\;$In particular,

\begin{equation}
\mu \left( p\right) _{orbit}=\frac{9M^{2}+\left( \zeta ^{2}-1\right) \tilde{m%
}^{2}}{3\left( M+\tilde{m}\right) \tilde{m}}a\left( \frac{e}{2M}\right)
\equiv \kappa _{orbit}\left( \frac{e}{2M}\right) .  \label{magmo-L}
\end{equation}
Adding up the components $\mu \left( B\right) =\mu \left( B\right) _{v}+\mu
\left( B\right) _{spin}+\mu \left( B\right) _{orbit}$ of Eqs.(\ref{magmo-v}%
), (\ref{magmo-sea}), and (\ref{magmo-L}), we have$\;\mu \left( p\right)
=\left( 1+\kappa _{spin}+\kappa _{orbit}\right) \left( \frac{e}{2M}\right)
.\;$The general result for octet baryon is 
\begin{equation}
\mu \left( B\right) =\left( 1+\kappa _{spin}+\kappa _{orbit}\right) \mu
\left( B\right) _{v}.
\end{equation}
This means that quark sea contributions can be absorbed by an overall
rescaling of quark magnetic moments. Because we are performing a flavor
SU(3) symmetric calculation, the magnetic moment D/F ratio is not altered.
Consequently all baryon moment {\em ratios} are unchanged from their SU(6)
limit values, {\em e.g.} $\mu _{p}/\mu _{n}=-3/2,\;etc$. This necessarily
requires that the quark sea modification be proportional to the original NQM
values.

For the principal enigma of why can the valence quarks alone yield a good
account of the magnetic moments, the $\chi QM$ offers a simple explanation:
the contributions from the orbital and spin angular momenta of the quark sea
have opposite signs, Eqs.(\ref{magmo-sea}) \& (\ref{magmo-L}): 
\begin{equation}
\kappa _{spin}=-\frac{7+2\zeta ^{2}}{3}a\;\;\;\;\;\;\;\;\;\;\;\kappa
_{orbit}=\frac{9M^{2}+\left( \zeta ^{2}-1\right) \tilde{m}^{2}}{3\left( M+%
\tilde{m}\right) \tilde{m}}a.
\end{equation}
This, of course, is intimately connected to the fact that the orbital and
spin alignments of the sea must be opposite to each other because of angular
momentum conservation, Eq.(\ref{angmo-cons}). In particular, for $\zeta $ in
the range of$\;\left( -1,0\right) ,$ we can have 
\begin{equation}
\kappa _{spin}+\kappa _{orbit}\simeq 0\text{ \ \ \ \ \ for\ \ \ }M\simeq 1.5%
\tilde{m}.  \label{cancel}
\end{equation}
The orbital contribution being dominated by the light GB processes, this
cancellation should be indicative of the actual situation. This diminution
means that even though $\Delta q_{v}$ is significantly different from $%
\Delta q_{\text{expt}},$ for a magnetic moment calculation we can still use $%
\Delta q_{v}$ if at the same time the orbital angular momentum contribution
is ignored. This explains why the NQM can give a satisfactory account of the
baryon magnetic moments even if its spin content prediction has been found
to be incomplete.

\underline{Remark-1}{\bf :} Previous discussions of the orbital angular
momentum contribution to the baryon magnetic moment\cite{confmx} have been
concerned with the configuration mixing, between the S-wave and possible
higher orbital states, of the three valence quarks rather than the
contribution by the orbital angular momentum of the quark sea. Our viewpoint
is that valence quark configuration mixing might not be a major factor
because the simple quark model is known to yield an adequate account of the
baryon magnetic moments. In a subsequent remark, we shall comment on the
issue of improving upon the NQM description.

\underline{Remark-2}{\bf : }Much of the current discussions on the proton
spin problem\cite{HYCheng} has to do with a possible gluonic contribution,
which is studied in terms of the Lagrangian (hence perturbative) degrees of
freedom --- in contrast to the nonperturbative QCD quantities of constituent
quarks and internal GBs of the present work. We view these two descriptions
as complimentary approaches: the validity of one does not preclude the
correctness of the other\cite{CLsch97}. An analogy with the baryon mass
problem is instructive. Even if one finds, via the energy-momentum trace
anomaly, that most of the baryon mass is gluonic in origin\cite{SVZ}, it is
still very useful to have the nonrelativistic QM picture of the baryon mass
being mostly the sum of its constituent quark masses. The additional mass of
a constituent quark results from QCD interactions, hence gluonic in origin.
(In $\chi $QM this gluonic interaction corresponds to the quark gaining a
large mass when propagating in the chiral condensate of the QCD vacuum.) In
the same manner, gluons can contribute to the baryon spin through the axial
anomaly. The analogy suggests the possibility of viewing the renormalization
effects due to the $\chi QM$ quark sea as ultimately corresponding to the
gluonic contribution.

\underline{Remark-3}{\bf : }The field theoretical calculation of the chiral
renormalization effects will be, to the leading order, that of the one-loop
diagrams with intermediates states of quarks and GBs. The relativistic
computation automatically includes both the sea quark-spin and orbital
angular momentum contributions. In fact, such $\chi $QM calculations have
been carried out\cite{Dicus}\cite{Brekke} and both groups found the
resultant anomalous magnetic moments of the constituent quarks to be small.
This lends support to our contention that there must be significant
cancellation among the spin and orbital angular momentum contributions.

\underline{Remark-4}{\bf : }Our present $\chi $QM discussion suggests that
to improve upon the NQM calculation of the baryon magnetic moments we can
start with the valence quarks, and augment them with the small anomalous
quark moments due to the chiral loop effects, and also include the
``exchange current effects'' due to the GB-exchanges among the valence quarks%
\cite{ex-curr}. Indeed, the study in Ref.\cite{Brekke} has concluded that
such a calculation does indeed yield a very satisfactory description of the
magnetic moments.

The conclusion we wish to draw is that the spin and magnetic moment data are
consistent with the $\chi $QM predictions: (A) a significantly polarized
quark sea in the direction opposite to the baryon spin, $\Delta q_{sea}<0$,
and yet (B) the antiquarks in the sea are not significantly polarized, $%
\left( \bar{q}_{+}-\bar{q}_{-}\right) =0,$ and (C) there should also be a
sizable amount of orbital angular momentum which because of conservation law
just cancels the quark polarization of the sea:$\;\left\langle
L_{Z}\right\rangle =-\frac{1}{2}$ $\Delta \Sigma _{sea}$. This diminishes
the quark sea contribution and allows for a successful description of the
baryon magnetic moments by the NQM.

One of us (L.F.L.) wishes to acknowledge the support by the U.S. Department
of Energy (Grant No. DOE-ER/40682-127).


\begin{references}
\bibitem{asym-mom}  With SU(3) breaking effects taken only in the quark
moments $\mu _{s}\simeq 2\mu _{d}/3$ or $M_{s}\simeq 3M_{d}/2$, and taking $%
\mu _{d}\simeq -0.9\,n.m.,$ we can obtain a fairly good fit of the measured
moments (to a general agreement of within $10\%,$ with the notable exception
of $\Xi ^{-}$ which is off by about $30\%$). This implies the constituent
quark masses of $M\simeq 350\,MeV$ and $M_{s}\simeq 525\,MeV.$ These values
just match, within expected uncertainty, those obtained in quark model
fittings (with spin-dependent contribution included) of the baryon masses,
see, for example, \cite{CLsch97} and references cited therein.
\end{references}
\end{document}